\documentclass[useAMS,iop,apj,numberedappendix,twocolappendix,usenatbib]{emulateapj}
\usepackage{amsmath}
\usepackage{amssymb}
\usepackage{graphicx}
\usepackage{color}
\usepackage{ae,aecompl}
\usepackage[pdfa]{hyperref}
\usepackage{multirow}

\slugcomment{}

\begin{document}
\title{On the gravitational stability of the Maclaurin disk}
\shorttitle{On the gravitational stability of the Maclaurin disk}
\author{Mahmood Roshan \altaffilmark{1,$*$} and Shahram Abbassi\altaffilmark{1,$\dagger$} and Habib G. Khosroshahi\altaffilmark{2,$\ddagger$}}
\affil{$^1$Department of Physics, Ferdowsi University of Mashhad, P.O. Box 1436, Mashhad, Iran}
\affil{$^2$School of Astronomy, Institute for Research in Fundamental Sciences (IPM), P.O. Box 19395-5531, Tehran, Iran}
\altaffiltext{$*$}{E-mail: mroshan@um.ac.ir}
\altaffiltext{$\dagger$}{E-mail: abbassi@um.ac.ir}
\altaffiltext{$\ddagger$}{E-mail: habib@ipm.ir}

%\documentclass[useAMS,usenatbib]{mnras}
%\usepackage{amsmath}
%\usepackage{amssymb}
%\usepackage{graphicx}
%\usepackage{txfonts}
%\usepackage{ae,aecompl}
%\title[On the gravitational stability of the Maclaurin disk]{On the gravitational stability of the Maclaurin disk}
%\author[first author et al.]{authors$^{1}$\thanks{E-mail:
%    email}\\ $^{1}$Department of Physics, School of Sciences, Ferdowsi University of Mashhad, Mashhad, 91775-1436, Iran}
%\begin{document}
%\def \aj {Astron. J.}
%\def \mnras {Mon. Not. R. Astron. Soc.}
%\def \apj {Astrophys. J.}
%\def \apjs {Astrophys. J. Suppl.}
%\def \apjl {Astrophys. J. Let.}
%\def \aap {Astron. \& Astrophys.}
%\def \aaps {Astron. \& Astrophys. Suppl. Ser.}
%\def \nat {Nature}
%\def \pasp {Pub. Astron. Soc. Pac.}
%\def \pasa {PASA}
%\def \araa {ARA\&A}
%\def \iaucirc {IAU\ Circ.}
%\def \prd {Phys.Rev.\ D}
%\def \pasj {PASJ}
%\label{firstpage}
%\maketitle}

\begin{abstract}
We study the global gravitational stability of a gaseous self-gravitating Maclaurin disk in the absence of a halo. Further, we replace Newtonian gravity with the specific Modified gravity theory known as MOG in the relevant literature. MOG is an alternative theory for addressing the dark matter problem without invoking exotic dark matter particles, and possesses two free parameters $\alpha$ and $\mu_0$ in the weak field limit. We derive the equilibrium gravitational potential of the Maclaurin disk in MOG and develop a semi-analytic method for studying the response of the disk to linear non-axysymmetric perturbations. The eigenvalue spectrum of the normal modes of the disk is obtained and its physical meaning has been explored. We show that Maclaurin disks are less stable in MOG than in Newtonian gravity. In fact both parameters $(\alpha,\mu_0)$ have destabilizing effects on the disk. Interestingly $\mu_0$ excites only the bar mode $m=2$ while $\alpha$ affects all the modes. More specifically, when $\alpha>1$, the bar mode is strongly unstable and unlike in Newtonian gravity can not be avoided, at least in the weak field limit, with increasing the pressure support of the disk.
\end{abstract}

\section{Introduction}
Maclaurin disk is a close fluid analog to the stellar Kalnajs disk, \citet{ka}. It is a two-dimensional thin disk of fluid, in which the pressure operates only in the plane of the disk and the surface density profile is the same as the Kalnajs disk. The horizontal modal properties of this disk can be investigated analytically, see \citet{ta1976} and \citet{we}, and for the vertical modes of the Kalnajs disk see \citet{polyachenko}. This means that there is an exact criterion for global gravitational stability of the disk. To the best of our knowledge this disk is the only self-gravitating fluid disk which its stability can be studied analytically. Although the disk is too far from a real galactic disk, its analytic structure is useful for understanding the global stability of more realistic galaxy models. In fact, as we discuss in what follows, there is a good agreement between simple analytic results of the Maclaurin disk and those obtained from numerical and N-body simulations of galactic disks.

The  most unstable mode in Maclaurin/Kalnajs disk is the $m=2$ mode, \citet{ta1976}, \citet{hu}, \citet{ka}. This mode is known as the bar mode. On the other hand N-body simulations of the galactic disks also revealed that bar mode is the most unstable mode not only for the Macluarin disk but for wide range of galaxy models. In fact, it has long been known from numerical simulations that rotationally supported stellar disks are globally unstable against the bar mode, for example see \citet{miller}, \citet{ho}, \citet{os}, \citet{se}, \citet{at1986}.

More importantly, the Maclaurin disk can be stabilized by embedding it in a rigid halo with constant matter density. In this case, one can again find an exact dispersion relation for propagation of the perturbations and the corresponding eigenvalue spectrum, see \citet{bt}. Rigid halos with non-uniform density, have also stabilizing effects on the gaseous disk. Using a semi-analytic method \citet{ta1978} investigated the stabilizing effects of non-uniform massive halos. This fact, namely the stabilizing role of a massive halo, is also a well-established feature of N-body and hydrodynamic simulations. \citet{os} using N-body simulations for studying the global stability of a galactic disk, realized that there should be massive halos around galaxies to avoid the rapid bar instability. For the first time in the literature, they created the "dark matter halo" concept. Criteria for stability against bar mode were formulated empirically by \citet{os} and \citet{ef}.

The overall outcome of the relevant literature in the past five decades is that the existence of a matter halo may stabilize the disk against global perturbations. More specifically the halo will reduce the growth rate of the stellar bar. It should be stressed that the incidence of bars in the real galactic disks is much larger than traditionally thought. In fact  \citet{sh} found in the COSMOS field that, in the local Universe, about 65\% of luminous spiral galaxies are barred. This fraction is a strong function of the cosmic redshift $z$, dropping to 20\% at a redshift of $z=0.8$. This means that stellar bars are effectively formed in the spiral galaxies. Furthermore bars are key structures that help to redistribute angular momentum between different components of disk galaxies, \citet{at2002}. Bars are thought to excite spiral arms \citep{to,san1976} and derive galactic ring patterns, for example see \citet{com} and \citet{ma}. They transport gas to the centers of disk galaxies and help to develop bulges and possibly trigger AGN activity.

Thus a question naturally arises that is it really necessary to consider dark matter halos in order to avoid stellar bars? 

In 1970-1980s, the halo was exploited to avoid the bar instability but now it is needed to slow down the bar growth. In other words, without the halo, the bar instability occurs and the bar grows in a timescale very short compared with the life time of spiral galaxies. This fact is in a gross disagreement with the bar fraction observations. Therefore, from this perspective, the problem is the growth rate of the bars and not their existence. Thus one may conclude that dark matter halo is still an important ingredient of disk galaxies and play key role in their evolution. 

We would like to stress that although existence of the dark matter halo provides a satisfactory picture for the bar evolution, it leaves also some serious puzzles. In some cases a halo can even stimulate the bar growth rather than restraining it, for example see \citet{at2002} and \citet{saha}. Recently \citet{se2016} using N-body simulations showed that angular momentum exchange between the live halo and the stellar disk is a source of bar instability. This means that if disk galaxies are embedded in live dark matter halos, then the observational fact that more than 35\% of them lack a strong bar is still a serious puzzle.

From our brief introduction, it is evident that there is a close relation between bar growth, bar fraction and the presence of dark matter halo. More specifically, dark matter halos not only stabilize the disk but also explain the flat rotation curves of the spiral galaxies, for a historical review of the dark matter problem we refer the reader to \citet{sand2010}. On the other hand the nature of the dark matter is still debated. This fact keeps open another serious approach to the problem: i.e. modified gravity which can offer some solutions to this problem. Theses theories are extensively used to find a way out from dark energy (for example see \citep{ca} for a review of dark energy models) and dark matter problem (for example see \citep{milgrom} and \citep{fa} for modified Newtonian dynamics (MOND) , and \citep{mo2006} for Modified Gravity (MOG), see section \ref{mog} for a brief introduction to MOG). 

The bar growth and the gravitational stability of astrophysical self-gravitating systems in modified gravity theories have a same importance as in standard gravity. More specifically, the observed bar fraction of the spiral galaxies should be consistent with the predictions of these theories without any need to dark matter halos. 

In the context of modified theories of gravity which have been presented to solve the dark matter problem, such as Tensor-Vector-Scalar theory (as a relativistic theory for MOND) \citet{bek} and MOG \citet{mo2006}, the cosmic structure formation should be promoted without cold dark matter particles. We remind that the structure formation is another form of gravitational instability known as Jeans instability. If a modified gravity cannot explain the observed spectrum for the cosmic inhomogeneities or cannot explain the observed growth rate and fraction of the stellar bars, then cannot also be considered as a viable gravity theory. Consequently, gravitational stability issues in the galactic and cosmological scales may provide a serious criterion for deciding about the viability of gravity theories. This fact makes our main motivation in this paper to study the stability of Maclaurin disk in MOG in order to shed light on the effects of MOG on the bar growth rate in real situations. In other words, regarding the practical role of the Maclaurin disk for better understanding of the global stability in Newtonian gravity, we have chosen it to compare the global stability of galactic disks in Newtonian gravity and MOG. It is necessary to mention that the local stability of spiral galaxies in MOG has been already investigated in \citet{ro2015}. 

In this paper we use a semi-analytic method to study the global stability of the Maclaurin disk. For other papers in which the modal properties of model galactic disks have been studied using semi-analytic methods see \citet{ev}; \citet{jal2005}; \citet{jal2007}. In \citet{chris},\citet{brada}, \citet{ti} and \citet{brand} the global stability of the galactic disks have been studied in different modified theories of gravity using N-body simulations. Naturally, N-body simulations can help to explore the bar growth in MOG in a more constructive way, Ghafourian \& Roshan (2016) in preparation.

\section{Modified Poisson equations in MOG}
\label{mog}
In this section we briefly review MOG and its main consequences in astrophysics. Since we study a non-relativistic Maclaurin disk, we also review the weak field limit of MOG and introduce the modified version of the Poisson equation.

MOG is a fully relativistic and covariant extension of GR. \citet{mo2006} introduced this theory to resolve the dark matter enigma. MOG is much more complicated than GR in the sense that its associated gravitational fields includes three types of fields while GR uses only a tensor field, i.e. the metric tensor. On the other hand, in MOG in addition to the metric tensor, there are two scalar fields ($\mu(x^{\beta})$ and $G(x^{\beta})$) and also a massive Proca vector field $\phi^{\beta}$. Naturally, these extra fields may help to handle the dark matter problem without invoking exotic dark matter particles. This theory has been successfully applied to explain the rotation curves of spiral galaxies and the mass problem in the galaxy clusters \citep{br2006,br2007,mo2006,mo2008,mo2009,mo2013a,mo2013,mo2014}. For cosmological consequences of this theory see \citet{mo2015}; \citet{roepjc} and \citet{Jamali}.

As we mentioned, MOG is a relativistic theory and we need its weak field limit for studying a non-relativistic Maclaurin disk. It should be stressed that real spiral galaxies are also in the weak field regime and the relativistic effects on their secular evolution are negligible. In this limit, MOG leads to two differential equation replaced with the standard Poisson equation. For the details of deriving the weak field limit of MOG, we refer the reader to \citet{mo2013}, \citet{ro2014}, \citet{ro2013}. The test particle's equation of motion takes the following form
\begin{equation}
\frac{d^2\mathbf{r}}{dt^2}=-\nabla \Phi
\end{equation}
where $\Phi$ defined as an effective gravitational potential and is given by 
\begin{equation}
\Phi=\Psi+ \phi
\label{pot1}
\end{equation}
where $\phi=\zeta \phi^{0}$ in which $\zeta$ is a coupling constant measuring the coupling strength of the vecor field to the ordinary matter and $\phi^0$ is the zeroth component of the Proca vector field. Furthermore, $\Psi$ and $\phi^0$ satisfy the following equations
\begin{equation}
\nabla^2\Psi=4\pi (1+\alpha) G \rho
\label{var}
\end{equation}
\begin{equation}
(\nabla^2-\mu_0^2)\phi=-4\pi\alpha G\rho
\label{vec}
\end{equation}
where $\mu_0$ and $\alpha$ are the free parameters of the theory in the weak filed limit, $G$ and $\rho$ are the gravitational constant and the matter density, respectively. In fact, $\mu_0$ is the background value of the scalar field $\mu$ and $\alpha$ is related to the coupling constant $\zeta$ as $\alpha=\zeta^2/\omega_0G$. Note that $\omega_0$ is another coupling constant. In fact there are two coupling constants $\zeta$ and $\omega_0$ in this theory. However, in the weak field limit these coupling constants always combine to form a single parameter $\alpha$. On the other hand the background value of $\mu$ appears as a new free parameter. For moe details see \citet{ro2014}. The observational values of the free parameters $\alpha$ and $\mu_0$ are known from rotation curve data of spiral galaxies. It has been shown by \citet{mo2013} that the best values for these parameters are $\alpha=8.89\pm 0.34$ and $\mu_0=0.042\pm 0.004 kpc^{-1}$. 

In order to study the global stability of the Maclaurin disk, in addition to the generalized Poisson equations, we need also the continuity and Euler equations in the context of MOG. Fortunately the mathematical form of these equations are the same as in Newtonian gravity, and it is just needed to replace the Newtonian gravitational potential with the effective potential $\Phi$, see \citet{ro2014} for details. Therefore we can write
\begin{equation}
\frac{\partial \rho}{\partial t}+\nabla \cdot (\rho \mathbf{v})=0
\label{conti}
\end{equation}
\begin{equation}
\frac{\partial \mathbf{v}}{\partial t}+(\mathbf{v}\cdot \nabla)\mathbf{v}=-\frac{\nabla p}{p}-\nabla \Phi
\label{euler}
\end{equation}
where $p$ is the pressure and $\mathbf{v}$ is the velocity of the fluid. Equations (\ref{var})- (\ref{euler}) combined with the equation of state of the fluid, make a complete set of equations for describing the dynamics of a self-gravitating fluid system in MOG. 

\section{Gravitational potential of Maclaurin disk in MOG }
\label{pot}
In this section we find the gravitational potential of a Maclaurin disk in the context of MOG. To do so we assume a non-rotating cylindrical system $(R,\varphi, z)$, such that $z$ axis coincide with the rotation axis of the disk and the angle $\varphi$ increases in the direction of rotation. We show that the potential can be obtained analytically. The Maclaurin disk is a fluid disk and its surface density for $R<a$ is given by 
\begin{equation}
\Sigma_0(R)=\Sigma_c\sqrt{1-\frac{R^2}{a^2}}
\label{density}
\end{equation}
and for $R>a$ the surface density is zero. Where $a$ is the radius of the disk edge and $\Sigma_c$ is a constant denoting the surface density at the center. The equation of state is $p=K \Sigma^3$. In Newtonian gravity the gravitational potential can be analytically obtained and, as we mentioned in the introduction section,  the global gravitational stability of the disk can be analytically studied. Let us start with the modified Poisson equations (\ref{var}) and (\ref{vec}). Equation (\ref{var}) is the standard Poisson equation and the only difference is the appearance of $(1+\alpha)G$ instead of $G$ in the right hand side. Therefore as in Newtonian gravity one may easily verify that the solution of this equation for the surface density (\ref{density}) is given by (see Biney \& Tremaine 2008)
\begin{equation}
\Psi(R)=(1+\alpha)a^2 \Omega_0^2\left[\frac{R^2}{2a^2}-1\right]
\end{equation} 
where 
\begin{equation}
\Omega_0^2=\frac{\pi^2 G \Sigma_c}{2 a}
\end{equation}
In order to solve equation (\ref{vec}) on the surface of the disk, let us use the separation of variables method as $\Phi(R,z)=F(R)Z(z)$. Substituting this expression into (\ref{vec}) we find two differential equations. By solving these differential equations for $z\neq 0$, i.e. in vacuum, one may easily verify that
$F(R)\propto J_0(\kappa R) $ and $Z(z)\propto e^{-k|z|}$. Where $J_0$ is the cylindrical Bessel function of order zero , $k$ is a real and positive constant and $\kappa$ is defined as $\kappa=\sqrt{k^2-\mu_0^2}$. Note that $k^2\geq\mu_0^2$, or equivalently Im$(\kappa)=0$, otherwise the modified Bessel function $I_0$ will appear in the solution instead of $J_0$. Since $I_0$ increases with radius, it will not satisfy all the conditions required for it to be the gravitational potential of an isolated surface density. 

 Therefore let us consider the potential
\begin{equation}
\phi_k(R,z)=e^{-k|z|}J_0(\kappa R)
\end{equation}
this potential solves equation (\ref{vec}) everywhere except in the plane $z=0$. Also one may easily show that since matter is located at $z=0$ (we recall that $\rho(R,z)=\Sigma(R)\delta(z)$), the derivative of $\phi_k$ with respect to $z$ is not continuous at $z=0$. In order to find the surface density $\Sigma_k(R)$ that generates the potential $\phi_k$, we integrate equation (\ref{vec}) with respect to $z$ in the interval $z=-\xi$ to $z=+\xi$, where $\xi$ is a positive constant,
and then let $\xi\rightarrow 0$. The result is
\begin{equation}
\Sigma_k(R)=\frac{k}{2\pi G \alpha}J_0(\kappa R)
\label{sk}
\end{equation}
Therefore for an arbitrary density we can write
\begin{equation}
\phi(R,z)=\int_{\mu_0}^{\infty}S(k)\phi_k(R,z)dk
\label{1}
\end{equation}
where $S(k)$ is related to the surface density through the following Bessel integral 
\begin{equation}
\Sigma(R)=\int_{\mu_0}^{\infty}S(k)\Sigma_k(R)dk
\end{equation}
$S(k)$ can be written as a function of $\kappa$
\begin{equation}
S(\kappa)=2\pi G\alpha\int_{0}^{\infty}\Sigma(R)J_0(\kappa R)R dR
\label{2}
\end{equation}
where we have used (\ref{sk}) and the fact that $k dk=\kappa d\kappa$. Potential on the disk can be obtained by using equations (\ref{1}) and (\ref{2})
\begin{equation}
\phi(R)=2\pi G\alpha \int_{0}^{\infty} \frac{J_0(\kappa R)\kappa}{\sqrt{\kappa^2+\mu_0^2}} d\kappa\int_{0}^{\infty} \Sigma(R')J_0(\kappa R')R' dR'\nonumber
\end{equation}
the second integral in the right hand side can be simply integrated for the Maclaurin disk. The result is
\begin{equation}
\phi(R)=2\pi G\alpha \Sigma_c \sqrt{\frac{\pi a}{2}}\int_{0}^{\infty} \frac{\kappa^{-\frac{1}{2}}d\kappa}{\sqrt{\kappa^2+\mu_0^2}}J_0(\kappa R) J_{\frac{3}{2}}(\kappa a)
\label{int1}
\end{equation}
This is a complicated integral to be solved exactly. In order to find an analytic solution, let us assume that $\mu_0 a\ll 1$. From modified Poisson equations (\ref{var}) and (\ref{vec}) it can be deduced that in the limit $\mu_0\rightarrow 0$, Newtonian gravity is recovered. Therefore it is natural to expect that $1/\mu_0$ is much larger than the characteristic length of the system, i.e. $a$. On the other hand, we recall that for real spiral galaxies $\mu_0\simeq 0.042 kpc^{-1}$ and by assuming $a\simeq 10 kpc$ as a typical value for the galactic disk radius, we find that $\mu_0 a\simeq 0.1$. Therefore even for real galaxies this assumption is somehow reasonable. However, here we are working on an idealized galaxy model in order to shed light on the effects of modified gravity on the global stability of the disk galaxies. Consequently, regarding our aim, this assumption is not restrictive and will simplify our analysis. 

Therefore, in order to solve integral (\ref{int1}) in the approximation $\mu_0 a\ll 1$, we differentiate it with respect to $R$ and expand it as power series in $\mu_0 a$, namely
\begin{eqnarray}
\frac{d\phi}{dR}&=&-2\pi G\alpha \Sigma_c \sqrt{\frac{\pi a}{2}}\int_{0}^{\infty} \frac{\kappa^{\frac{1}{2}}d\kappa}{\sqrt{\kappa^2+\mu_0^2}}J_1(\kappa R) J_{\frac{3}{2}}(\kappa a)\nonumber\\ &\simeq&-2\pi G\alpha \Sigma_c \sqrt{\frac{\pi a}{2}}\int_{0}^{\infty} \kappa^{-\frac{1}{2}}J_1(\kappa R) J_{\frac{3}{2}}(\kappa a) dk \\ &+& \mu_0^2a^2 \pi G\alpha \Sigma_c \sqrt{\frac{\pi}{2a^3}}\int_{0}^{\infty} \kappa^{-\frac{5}{2}}J_1(\kappa R) J_{\frac{3}{2}}(\kappa a) dk \nonumber\\&+&O (\mu_0^3a^3)\nonumber
\end{eqnarray}
where we have truncated the series at terms of order $\mu_0^2a^2$. These integrals can be exactly solved, see page 683 in \citet{gr}. The result is
\begin{equation}
\frac{d\phi}{dR}=\alpha \Omega_0^2 \left(\left(\frac{1}{4}a^2\mu_0^2-1\right)R-\frac{\mu_0^2}{16}R^3\right)+O(\mu_0^3a^3)
\end{equation}   
Now this equation can be trivially integrated to obtain $\phi(R)$. Combining the result with (\ref{pot1}), the total potential $\Phi(R)$ of the disk for $R<a$ is 
\begin{equation}
\Phi(R)=\Omega_0^2a^2\left(\frac{R^2}{2a^2}-1\right)+ \frac{\mu_0^2 a^2\alpha \Omega_0^2}{8} \left(1-\frac{R^2}{8 a^2}\right)R^2+c
\label{pot2}
\end{equation}
where $c$ is an integration constant. The first term in the right hand side is the standard Newtonian potential and the second term is the corrections induced by MOG. One may easily verify that this potential leads to a stronger force than Newtonian gravity. This is a common feature among the theories which try to address the dark matter problem without invoking non-baryonic particles. It is also important mentioning that in contrast to Newtonian gravity, potential (\ref{pot2}) leads to a differentially rotating disk. In other words the angular velocity $\Omega(R)$ is no longer constant in MOG. This fact induces some difficulties for finding the global normal modes of the Maclaurin disk. The angular velocity is
\begin{eqnarray}
\Omega(R)^2=\frac{1}{R}\frac{d\Phi}{d R}+\frac{1}{R\Sigma}\frac{dp}{d R}=\Omega_N^2\\ \nonumber+\frac{\mu_0^2 a^2\alpha\Omega_0^2}{4}\left(1-\frac{R^2}{4 a^2}\right)
\label{av}
\end{eqnarray}
where $\Omega_N$ is the angular velocity of the disk in Newtonian gravity and is given by
\begin{equation}
\Omega_N^2=\Omega_0^2-\frac{3 K \Sigma_c^2}{a^2}
\end{equation}
\section{linear perturbation analysis}
\label{lin}
In this section we study the propagation of the linear horizontal perturbations on the disk. As we mentioned before, the Euler and continuity equations in MOG are the same as in Newtonian dynamics. After linearising in the cylindrical coordinate system, these equations can be written as (Binney \& Tremaine 2008)
\begin{equation}
\frac{\partial \Sigma_{1}}{\partial t}+\frac{1}{R}\frac{\partial}{\partial R}\left(\Sigma_{0} R v_{R1}\right)+\Omega \frac{\partial \Sigma_{1}}{\partial \varphi}+\frac{\Sigma_{0}}{R} \frac{\partial v_{\varphi 1}}{\partial \varphi}=0
\label{contin2}
\end{equation}
\begin{equation}
\frac{\partial v_{R1}}{\partial t}+\Omega \frac{\partial v_{R1}}{\partial \varphi}-2\Omega v_{\varphi 1}=-\frac{\partial}{\partial R}\left(\Phi_{1}+h_{1}\right)
\label{eulerr2}
\end{equation}
\begin{equation}
\frac{\partial v_{\varphi 1}}{\partial t}+\Omega \frac{\partial v_{\varphi 1}}{\partial \varphi}+\frac{\kappa^{2}}{2\Omega} v_{R1}=-\frac{1}{R}\frac{\partial}{\partial \varphi}\left(\Phi_{1}+h_{1}\right)
\label{eulerphi2}
\end{equation}
where $v_R$ and $v_{\varphi}$ are the radial and azimuthal velocities respectively, the subscript $"0"$ refers to the background value of the given quantity and $"1"$ denotes the
corresponding perturbed quantity. Also $h$ is the specific enthalpy defined as $h= \int dp/\Sigma$ and the epicyclic frequency $\kappa$ is defined as
\begin{equation}
\kappa(R)=\sqrt{R \frac{d\Omega^{2}}{dR}+4 \Omega^{2}}
\label{epip}
\end{equation}
using the equation of state of the disk, i.e. $p=K \Sigma^3$, one can show that $h_1=c_s^2\Sigma_1/\Sigma$, where $c_s$ is the sound speed. It turns out that using the oblate spheroidal coordinate system is more appropriate than the cylindrical coordinate system for studying the linear perturbations. The oblate spheroidal coordinates $\eta$, $\xi$ are defined as, for more details see \citet{hu},
\begin{eqnarray}\label{coo}
x&=&a\sqrt{(1+\xi^2)(1-\eta^2)}\cos\varphi\\\nonumber
y&=&a\sqrt{(1+\xi^2)(1-\eta^2)}\sin\varphi\\\nonumber
z&=&a \xi \eta
\end{eqnarray}
In this coordinate system plane of the disk is specified by $\xi=0$. Therefore we have $\eta=\sqrt{1-R^2/a^2}$.

Now let us assume that the perturbations can be written as a Fourier mode $Q=Q(\eta) e^{i (m \varphi-\omega t)}$. Also in the new coordinate system, we rewrite equations (\ref{contin2})-(\ref{eulerphi2}) with respect to dimensionless variables as
\begin{equation}
i(\omega-m \Omega)\eta\sigma_1+\frac{\partial}{\partial \eta}(\eta\sqrt{1-\eta^2} u_1)-\frac{i m \eta^2}{\sqrt{1-\eta^2}}v_1=0
\label{per1}
\end{equation}
\begin{equation}
i(\omega-m \Omega)u_1+2\Omega v_1=-\frac{\sqrt{1-\eta^2}}{\eta}\frac{\partial}{\partial\eta}(\Phi_1+\beta \eta \sigma_1)
\label{per2}
\end{equation}
\begin{equation}
i(\omega-m \Omega)v_1-k(\eta)u_1=\frac{i m}{\sqrt{1-\eta^2}}(\Phi_1+\beta \eta \sigma_1)
\label{pert3}
\end{equation}
where $\beta=\frac{3K\Sigma_c^2}{a^2\Omega_N^2}=(\frac{\pi^2 G a}{6 K \Sigma_c}-1)^{-1}$ and  other dimensionless quantities are defined as
\begin{eqnarray}
\omega &\equiv &\frac{\omega}{\Omega_{N}},\chi=\frac{a^2\mu_0^2\alpha}{8}(1+\beta),  \sigma_1=\frac{\Sigma_1}{\Sigma_c},\\\nonumber u_1&=&\frac{v_{R1}}{a\Omega_N}, v_1=\frac{v_{\varphi 1}}{a \Omega_N}, \Phi_1\equiv\frac{\Phi_1}{a^2\Omega_N^2}\\ \nonumber
k(\eta)&=&\frac{\kappa^2}{2\Omega\Omega_{N}}\simeq (2+\chi)+\chi \eta^2\\ \nonumber
\Omega & \equiv &\frac{\Omega}{\Omega_{N}}\simeq \Big(1+\frac{3\chi}{4}\Big)+\frac{\chi}{4}\eta^2
\end{eqnarray}
Equations (\ref{per1})-(\ref{pert3}) are the main equations of this section. In the following we will solve them semi-analytically to study the global stability of the disk in the context of MOG. Therefore it seems that we have three equations for four unknowns $u_1$, $v_1$, $\sigma_1$ and $\Phi_1$. However, we still have two more equations (\ref{var}) and (\ref{vec}) which relate the surface density perturbation $\sigma_1$ to the corresponding gravitational potential $\Phi_1$. The modified Poisson equations can be exactly solved in the spheroidal coordinate. We have done this in the Appendix and the final results can be summarized in the equations (\ref{dj}) and (\ref{fpot}). Theses equations are general solutions for non-axisymmetric normal modes $m\neq 0$. In fact solutions are oblate spheroidal wave functions. On the other hand spheroidal wave functions can be expanded in terms of the associate Legendre functions $P_{n}^m(\eta)$. We use these orthogonal eigenfunctions to expand the other perturbations. Using the prescription presented in \citet{ta1978} and \citet{hu}, one may expand the perturbations as
\begin{eqnarray}\label{dper}
\sigma_1 &=& \frac{1}{\eta}\sum_{l=0}^{\infty}\frac{A_l}{c_l} P_{m+2l}^m(\eta)e^{i (m \varphi-\omega t)}\\ \nonumber
\Phi_1 &=&\sum_{l=0}^{\infty}[(A_l+B_l) P_{m+2l}^m(\eta)+a^2\mu_0^2 [\psi_{m,m+2l}P_{m+2l-2}^m(\eta)\\ \nonumber ~~&+&\psi_{m,m+2l}'P_{m+2l+2}^m(\eta)]e^{i (m \varphi-\omega t)}B_l\\ \nonumber
u_1&=&\frac{i}{\sqrt{1-\eta^2}}\sum_{l=0}^{\infty}a_l P_{m+2l}^{m}(\eta)e^{i (m \varphi-\omega t)}\\ \nonumber
v_1&=&\frac{1}{\sqrt{1-\eta^2}}\sum_{l=0}^{\infty}b_l P_{m+2l}^{m}(\eta)e^{i (m \varphi-\omega t)}
\end{eqnarray}
Where we have used equations (\ref{dj}) and (\ref{fpot}). We mention that $A_l$ and $B_l$ are different from those appeared in the Appendix. In fact we have divided $A_l$ and $B_l$ of the Appendix by $\Omega_N^2 a^2$ to make them dimensionless. Also $c_l$ is defined as
\begin{eqnarray}
c_l=-(1+\alpha)(1+\beta)g_{m,m+2l}
\end{eqnarray}
we recall that for $l,m\neq 0$ the above perturbation conserve the mass of the disk. In other words $\int \sigma_1 ds=0$ where $ds$ is the surface element. Let us first recover the stability criterion in Newtonian gravity by setting $\alpha$ and $\mu_0$ to zero. In this case, it is clear from equation (\ref{rel}) that $B_l=0$. Also $\Omega(\eta)=1$ and $k(\eta)=2$. Therefore, after some algebraic manipulations, equations (\ref{per1})-(\ref{pert3}) can be combined to obtain 
\begin{equation}
\sum_{l=m}^{\infty}{\frac {iT_{{{\it lm}}}A_l P_{l}^{m}(\eta)}{ \left( -4+{\omega_{{r}}}^{2} \right) 
 \left( 1+\beta \right) g_{{{\it lm}}}}}=0
 \label{n1}
\end{equation}
where 
\begin{eqnarray}\label{ex}
T_{{{\it lm}}} ={\omega_{{r}}}^{3}+ \left(  \left( l-{m}^{2}+{l}^{2}
 \right)  ( 1+\beta \right) g_{{{\it lm}}}-4\\ \nonumber - \left( {l}^{2}+l-{
m}^{2} \right) \beta ) \omega_{{r}}-2\, \left(  \left( 1+\beta
 \right) g_{{{\it lm}}}-\beta \right) m
\end{eqnarray}
where $\omega_r=\omega-m$. From (\ref{n1}) one may conclude that $T_{lm}=0$. This is the well-known dispersion relation for the Maclaurin disk in Newtonian gravity, see \citet{bt}. It turns out that, in our notation, the stability criterion can be simply expressed as $\beta>1$. It is also well-established that the $m=2$ mode is the  most unstable mode. This mode is known as the bar mode. 

As we mentioned before, although this disk is very far from a real galaxy, its structure and normal modes, may help to better understanding of the results of more realistic N-body galaxy models. For example eigenvalue spectrum of the disk shows that the most unstable mode is the bar mode $m=2$. This is also what is seen in the simulation of the disk galaxies.

However, in MOG the stability criterion can not be obtained by combining the first order equations. In other words, one can not find a well behaved dispersion relation for propagation of the normal modes on the disk. In the following we use a different procedure to investigate the normal modes. In fact we reduce the problem to an eigenvalue problem and find the eigenvalue spectrum using a standard numerical procedure. 

Inserting equations (\ref{dper}) into linear equations (\ref{per1})-(\ref{pert3}), and using the orthogonality relations among the associated Legendre functions, we find the following equations
\begin{equation}
\sum_{l=0}^{\infty}c_k\Big[ \frac{3\chi}{4c_l}E_{kl}A_l-(F_{kl}+\delta_{kl})a_l+m G_{kl}b
_l\Big]=\omega_r^* A_k
\label{mat1}
\end{equation}
\begin{eqnarray}\label{mat2}
\sum_{l=0}^{\infty}\Big[\Big(\frac{\beta}{c_l}+1\Big)C_{kl}A_l+W_{kl}A_l+\frac{m \chi}{4} E_{kl}a_l\\ \nonumber+\frac{\chi}{2} E_{kl}b_l+\Big(2+\frac{3\chi}{2}\Big)\delta_{kl}b_l\Big]=\omega_r^* a_k
\end{eqnarray}
\begin{eqnarray}\label{mat3}
\sum_{l=0}^{\infty}\Big[m Z_{kl}A_l+((2+\chi)\delta_{kl}+\chi E_{kl})a_l\\ \nonumber+\frac{m\chi}{4}E_{kl}b_l\Big]=\omega_r^* b_k
\end{eqnarray}
where $\delta_{kl}$ stands for Kroneker's delta function, $\omega_r^*$ is related to $\omega_r$ as $\omega_r^*=\omega_r-3m\chi/4$ and the coefficients are defined as 
\begin{eqnarray}
E_{kl}&=&\frac{1}{\theta_k}\int_0^1 \eta^2 P_{m+2l}^m(\eta) P_{m+2k}^m(\eta) d\eta\\ \nonumber
F_{kl}&=&\frac{1}{\theta_k}\int_0^1 \eta \frac{dP_{m+2l}^m(\eta)}{d\eta} P_{m+2k}^m(\eta) d\eta\\\nonumber
G_{kl}&=&\frac{1}{\theta_k}\int_0^1 \frac{\eta^2}{1-\eta^2} P_{m+2l}^m(\eta) P_{m+2k}^m(\eta) d\eta\\\nonumber
C_{kl}&=&\frac{1}{\theta_k}\int_0^1 \frac{1-\eta^2}{\eta} \frac{dP_{m+2l}^m(\eta)}{d\eta} P_{m+2k}^m(\eta) d\eta\\\nonumber
\end{eqnarray}
and
\begin{eqnarray}
W_{kl}=\frac{a^2\mu_0^2}{\theta_k}\int_0^1 \frac{1-\eta^2}{\eta}\Big\lbrace \zeta_{ml}\frac{dP_{m+2l-2}^m(\eta)}{d\eta}~~~~~~~~~~~~~~\\\nonumber\zeta_{ml}' \frac{dP_{m+2l+2}^m(\eta)}{d\eta}\Big\rbrace P_{m+2k}^m(\eta)d\eta~~~~~~~~~~~~~\\\nonumber
Z_{kl}=\Big(\frac{\beta}{c_l}+1\Big)\delta_{kl}+a^2\mu_0^2\Big\lbrace\Big[\frac{\psi_{m,m+2l}}{c_l}-\frac{\epsilon_{l+1}'}{\alpha_{l+1}\alpha_l}\Big]\\\nonumber
\delta_{k,l+1}\Big[\frac{\psi_{m,m+2l}'}{c_l}-\frac{\epsilon_{l-1}}{\alpha_{l-1}\alpha_l}\Big]\delta_{k,l-1}\Big\rbrace~~~~~~~~~~~
\end{eqnarray}
where
\begin{eqnarray}
\theta_k &=&\frac{(2m+2k)!}{(2m+4k+1)2k!}\\\nonumber
\zeta_{ml}&=&\Big[\frac{\psi_{m,m+2l}}{c_l}-\frac{\epsilon_{l-1}}{\alpha_{l-1}\alpha_l}\Big]\\\nonumber
\zeta_{ml}'&=&\Big[\frac{\psi_{m,m+2l}'}{c_l}-\frac{\epsilon_{l+1}'}{\alpha_{l+1}\alpha_l}\Big]
\end{eqnarray}
and $\alpha_k$, $\epsilon_k$ and $\epsilon_k'$ have been defined in (\ref{ak}). Now let us rewrite equations (\ref{mat1})-(\ref{mat3}) in the matrix form $\mathbf{B}|\mathbf{r}>=\omega_r^* |\mathbf{r}>$ where
\[ \mathbf{B}=\left( \begin{array}{ccc}
\frac{3\chi c}{4}\frac{E}{c} & -c(F+\delta) & m c G \\
(\frac{\beta}{c}+1)C+W & \frac{m\chi}{4}E & \frac{\chi}{2}E+(2+\frac{3\chi}{2})\delta \\
m Z & (2+\chi)\delta+\chi E & \frac{m\chi E}{4} \end{array} \right)\] 
and
\[|\mathbf{r}>=\left(\begin{array}{c}
A\\
a \\
b  \end{array} \right)\]
Therefore the problem has been reduced to an infinite dimensional eigenvalue problem. However in practice we start from a small and finite dimension and increase the dimension until complex eigenfrequencies converge. We mention that matrix $\mathbf{B}$ is a real and non-symmetric matrix. Therefore, it will have imaginary eigenvalues as well as real ones. Let us write the eigenvalue $\omega$ as $\omega_r=\omega_R+ i\omega_I$. Real eigenvalues correspond to rotating modes in the same or opposite direction as the disk. On the other hand complex eigenvalues correspond to growing or damping modes. Growing modes for which $\omega_R\neq 0$ are known are \textit{overstabilities}.

\section{Eigenvalue Spectrum}
\label{eig}
\begin{figure}
\centerline{\includegraphics[width=9cm]{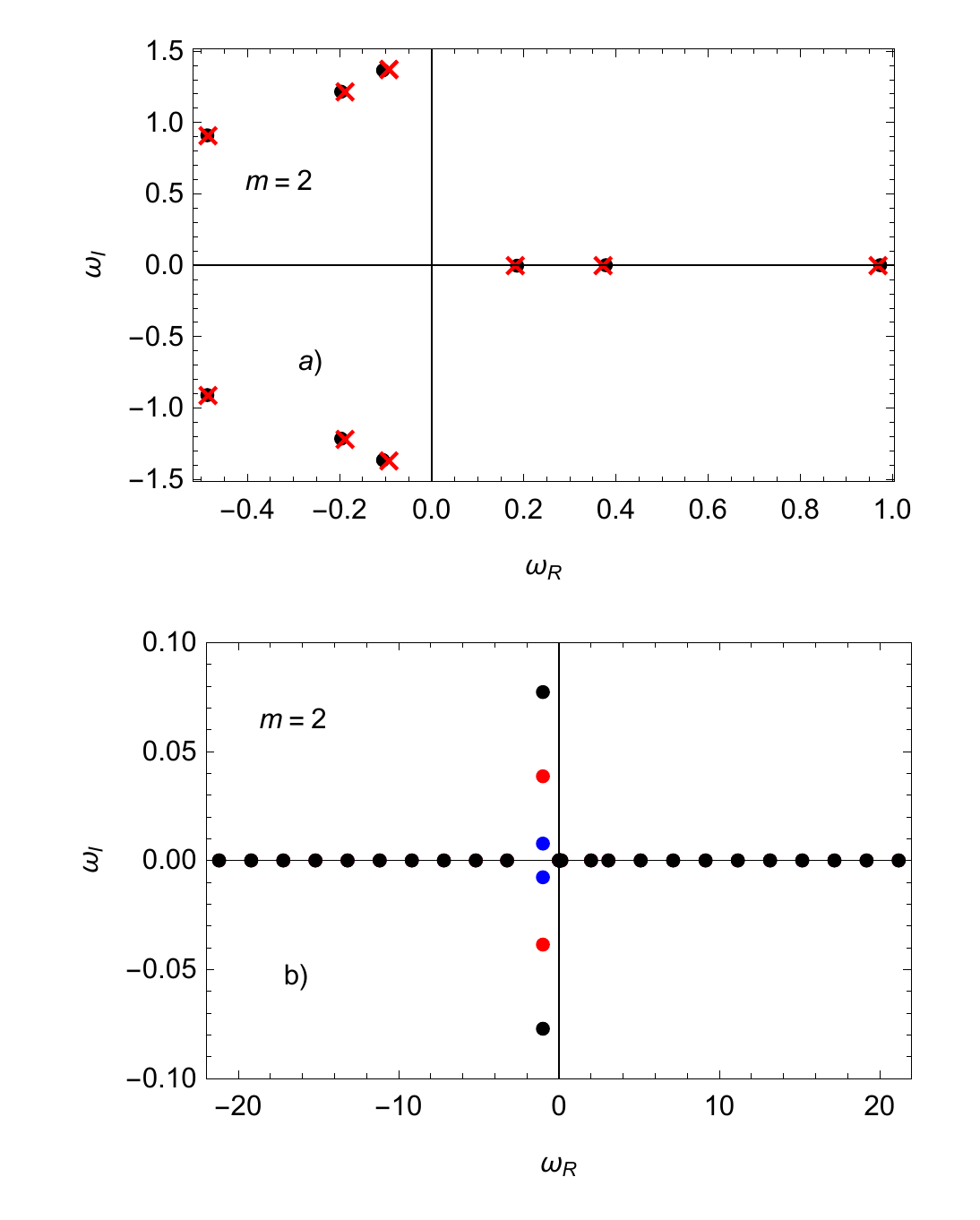}}
\caption[]{Fig. a) Comparison between eigenvalue spectrum obtained from the exact dispersion relation $T_{lm}=0$ and that obtained form the eigenvalue problem $\mathbf{B}|\mathbf{r}>=\omega_r^* |\mathbf{r}>$ (in Newtonian gravity). Red crosses correspond to a small part of the exact spectrum and the black points correspond to the eigenvalues of the matrix $\mathbf{B}$. In this figure $m=2$ and $\beta=0.08$. Fig. b) Blue, red and black circles correspond to $\mu_0=0.01$, $\mu_0=0.05$ and $\mu_0=0.1$ respectively. In this figure $\beta=1$. It is clear that by increasing $\mu_0$, the growth rate for the mode $(2,2)$ increases.}
\label{fig1}
\end{figure}

In the case of Newtonian gravity, where $\alpha$ and $\mu_0$ are zero, we expect that the above mentioned eigenvalue problem recovers the exact frequency spectrum obtained from equation (\ref{ex}). In Fig. \ref{fig1}a), we have shown a part of the exact eigenvalue spectrum for $m=2$ obtained from (\ref{ex}) by red crosses and those obtained from $\mathbf{B}|\mathbf{r}>=\omega_r^* |\mathbf{r}>$ with black points. In this figure we have shown $9$ points corresponding to different modes $(l,m)$. Also we assume that $\beta=0.08$, as we mentioned before for this choice for $\beta$ there are many unstable modes. It is clear that there is a good agreement between these methods. We have also applied this test for different modes $m$ in order to get sure that the numerical method leads to reliable results. 

Now let us consider the eigenvalues in the context of MOG. We recall again that $\beta$ can be considered as a stability parameter. In Newtonian gravity if $\beta>1$ then all $(l,m)$ modes are stable. However, in MOG we have two free parameters $(\alpha,\mu_0)$ and our main goal in this section is to study their effect on the stability of the disk. To do so, it is just enough to put different values of MOG's free parameters in the eigenvalue system $\mathbf{B}|\mathbf{r}>=\omega_r^* |\mathbf{r}>$ and interpret the spectrum.

 Let us start from the Yukawa mass parameter $\mu_0$. As we shall show, the system is not too sensitive to the changes in $\mu_0$. However, increasing this parameter excites some instabilities and increases the growth rate. To see this behavior more clearly, for the bar mode $m=2$ we set $\beta=1$ and $\alpha=0$, then find the spectrum for different values of $\mu_0$. Also it is necessary to mention that by increasing the dimension of the matrix $\mathbf{B}$ we realized that there is a suitable convergence when $\mathbf{B}$ is a $33\times 33$ matrix. Here we explain the criterion based on which we evaluate the convergence of the solutions. In fact we start with $n=3$. In this case $\mathbf{B}$  is a $12\times 12$ matrix and possesses $12$ eigenvalues corresponding to the first $12$ modes. Let us show these eigenvalues by $\omega_n^{(i)}$ where $i=1$ to $12$. By increasing the dimension, i.e $n$, of the stability matrix the eigenvalues $\omega_n^{(i)}$ will change. We increase $n$ and in each step we measure the following fractional difference
\begin{equation}
\frac{|\omega_n^{(i)}|-|\omega_{n+1}^{(i)}|}{|\omega_n^{(i)}|}
\end{equation} 
for all first $12$ modes of the system. We stop increasing $n$ when this fractional ratio for all modes is smaller than $10^{-6}$ and does not vary significantly with $n$. Of course this can be done for the real and imaginary parts of the eigenvalues as well. However the result will not affected. 
 
The result has been shown in Fig. \ref{fig1}b). It is clear from this figure that the disk undergoes more instabilities by increasing the $\mu_0$ parameter. More specifically $\omega_I$ grows linearly with $\mu_0$ as $\omega_I\simeq 0.78 \mu_0$. Albeit as in Newtonian gravity the instability can be avoided by increasing the magnitude of $\beta$. This means that we need to increase the pressure of the disk. This fact make sense since we know that MOG enhances the strength of gravity. Therefore more pressure support is needed to confront the gravitational force. It is also interesting to mention that, in this case $\mu_0\neq 0$ and $\alpha=0$, only the bar mode $m=2$ is excited. Although it is somehow reasonable since this mode is the most unstable mode in Newtonian gravity, this fact is clearly against what we expected from MOG. We recall that we expect that MOG behaves like dark matter halos and stabilize the galactic disks. However, we need to study the response of the system to the second parameter, namely $\alpha$, and then conclude about the effect of MOG on the global stability of the disk.
\begin{figure}
\centerline{\includegraphics[width=9cm]{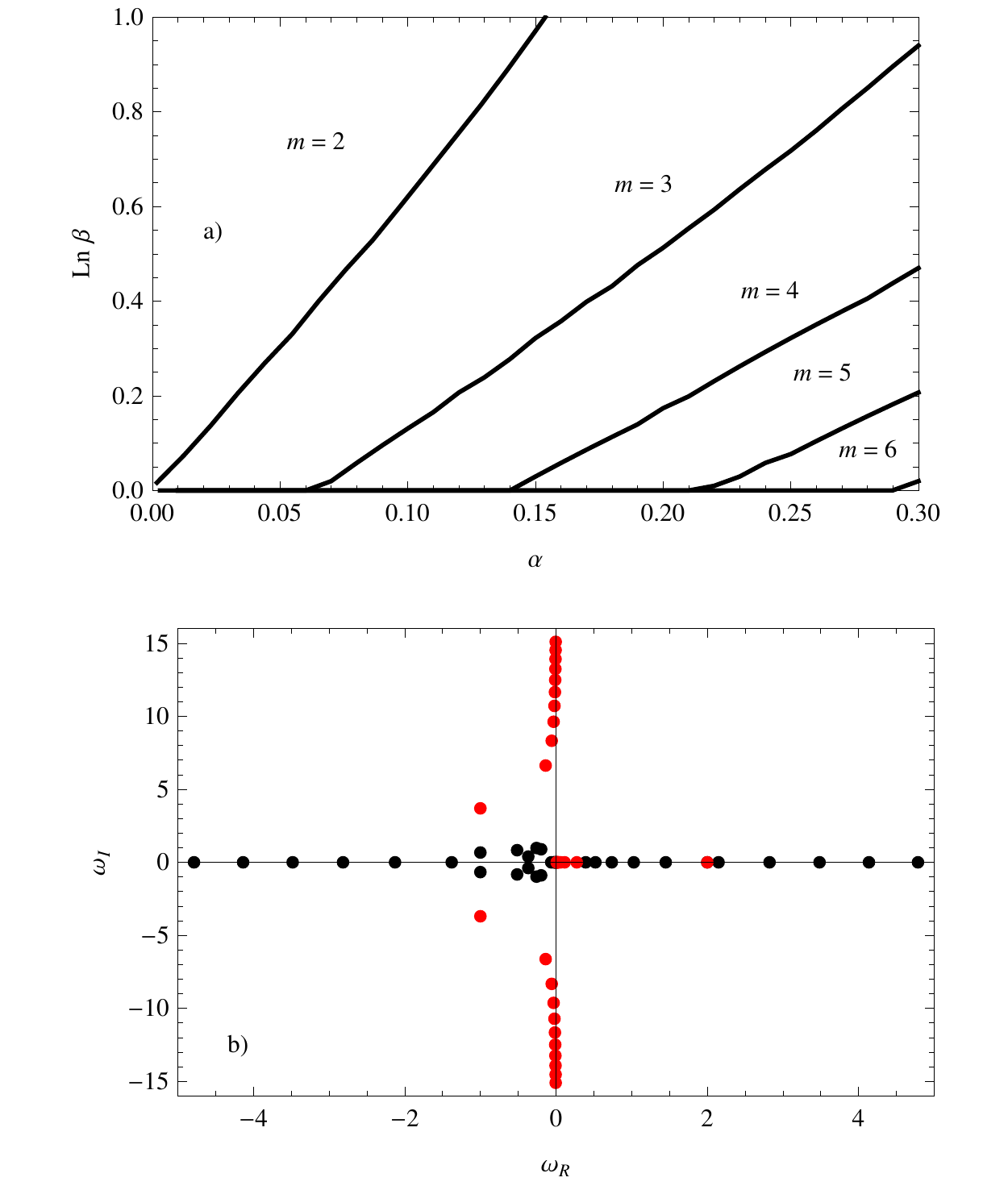}}
\caption[]{a) The border of stability of different modes has been shown for different values of $\alpha$ when $\mu_0=0.01$. In other words, the required value of $\beta$ for stabilizing the given $(m,\alpha)$ has been shown. b) Black points correspond to $m=2$ modes, with different $l$, in Newtonian gravity with $\beta=0.1$ and $\alpha=\mu_0=0$. Red points are the corresponding eigenvalues in MOG with $\alpha=8$ and $\mu_0=0.01$.}
\label{fig2}
\end{figure}

In Fig. \ref{fig2}a) we have shown the response of the system to non-axisymmetric perturbations when $\alpha\neq 0$. The mass parameter $\mu_0$ has been set to $0.01$ for all modes in this figure. In fact, in this figure we have plotted the border of the stability as curves $\beta(\alpha)$. For each $\alpha$, $\beta(\alpha)$ denotes a value for $\beta$ which makes the disk stable against the given perturbation. As one may expect, $\beta>1$ for all curves of Fig. \ref{fig2}a); and more pressure supports relative to Newtonian case is needed for stability. Interestingly, for $\alpha<0.06$ there is only one unstable mode, i.e. $m=2$. However increasing the $\alpha$ parameter other $m>2$ modes get unstable one by one.  

For the bar mode $m=2$ when $\alpha$ reaches $1$, required value of $\beta$ for stability reaches infinity. This means that in this case it is not possible to stabilize the disk by increasing the pressure. This is the case also for other modes. Albeit for $m>2$, $\beta$ reaches infinity at some larger values of $\alpha$. For example for $m=6$ the infinity happens at $\alpha\simeq 1.25$. This situation is reminiscent of the stability of the stellar Maclaurin disk in Newtonian gravity. The stellar version of the Maclaurin disk is known as the Kalnajs disk. In Newtonian gravity, increasing the pressure support of the system does not suppress the instabilities, and consequently all Kalnajs disks are unstable to at least one mode, see Fig. 5.5 in \citet{bt} for more details. 

On the other hand for Maclaurin disk in MOG, when $\alpha>1$ there will be at least one unstable mode which can not be avoided by increasing the pressure. Before discussing the observational values of $\alpha$ it is worth mentioning that at very large pressures, i.e. $\beta\gg 1$, the weak field limit of MOG is not reliable. In fact pressure may behave as a source of gravity in the relativistic situations. For example in the early universe where the dynamics of the cosmos is governed with the relativistic matter, the expansion rate of the universe is smaller than the matter dominated universe where the dynamics is determined with non-relativistic matter. More specifically the cosmic scale factor $a(t)$ grows as $t^{1/2}$ in the radiation dominated phase and as $t^{2/3}$ in the matter dominated universe. In other words, pressure in the early universe behaves like gravity and slows down the expansion rate. Therefore, in principle, fluid pressure may help the gravitational instability. In this case, it is necessary to add the post-Newtonian corrections to the hydrodynamic equations, \citet{po}. However, in this paper we have restricted ourselves to a non-relativistic disk.

Therefore it is important emphasizing that one can not claim that increasing the pressure support can not avoid the instability. In other words, this behavior, i.e. $\beta\rightarrow \infty$ when $\alpha\rightarrow 1$ for instance for $m=2$, is due to the limitations we have imposed on the field equations. More specifically the main equations \eqref{var}-\eqref{euler} are valid only in the weak field limit where $\beta$ is not too large. In fact, using the definition of $\beta$, the constraint $\beta \gg 1$ can be written as
\begin{equation}\label{n1}
\frac{p}{\Sigma\, c^2}\gg\Big( \frac{a \Omega_{N}}{c}\Big)^2
\end{equation}
On the other hand, it is convenient to assume that $a^2 \Omega_{N}^2=v^2\sim \Phi$, where $\Phi$ is the gravitational potential. In this case equation \eqref{n1} can be written as $p/\Sigma \gg \Phi$. This expression explicitly means that one can not ignore the gravitational effects of the fluid pressure, and the governing equations should be modified in order to take into account the general relativistic effects of the high pressure. 
\begin{figure*}
\centerline{\includegraphics[width=14cm]{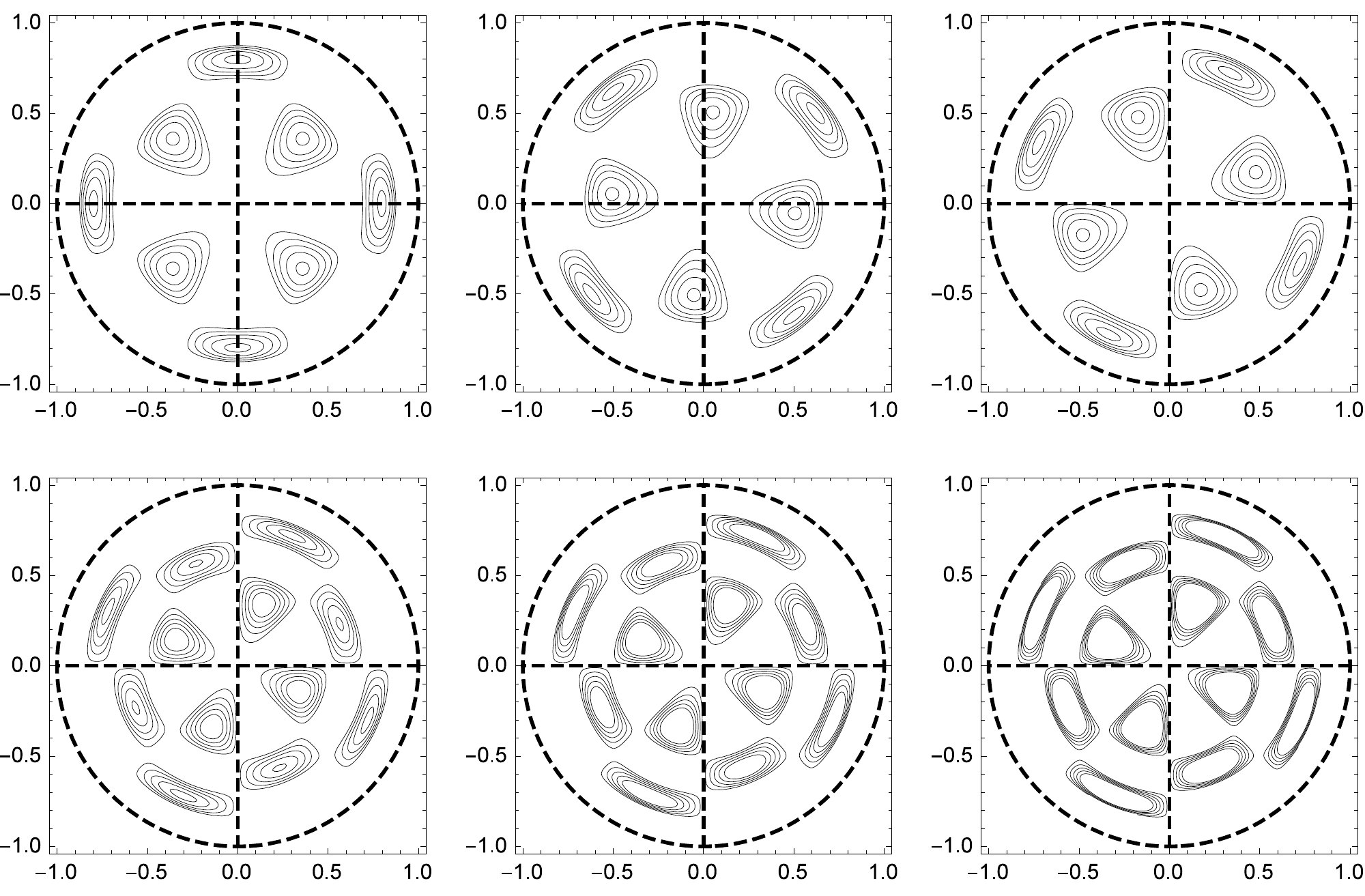}}
\caption{The contour plots show the real part of $\sigma_1$ for a $m=4$ mode. The contour levels range from 10\% to 90\% of the maximum of $\sigma_1$ with increments of 10\%. The top three plots belong to the fifth eigenmode of $m=4$ in Newtonia gravity with $\omega=-9.07$ and $\beta=1.1$. From left to right these three plots are in the time $t=0, 0.3,0.5$ respectively. The bottomn three plots belong to the same mode in MOG with $\alpha=8$, $\mu_0=0.01$ and $\beta=1.1$. In this case the eigenvalue is $\omega=0.023+ 10.82i$. From left to right these three plots are in the time $t=0, 0.03,0.07$ respectively. \label{fig3}}
\end{figure*}
\begin{figure*}
\centerline{\includegraphics[width=14cm]{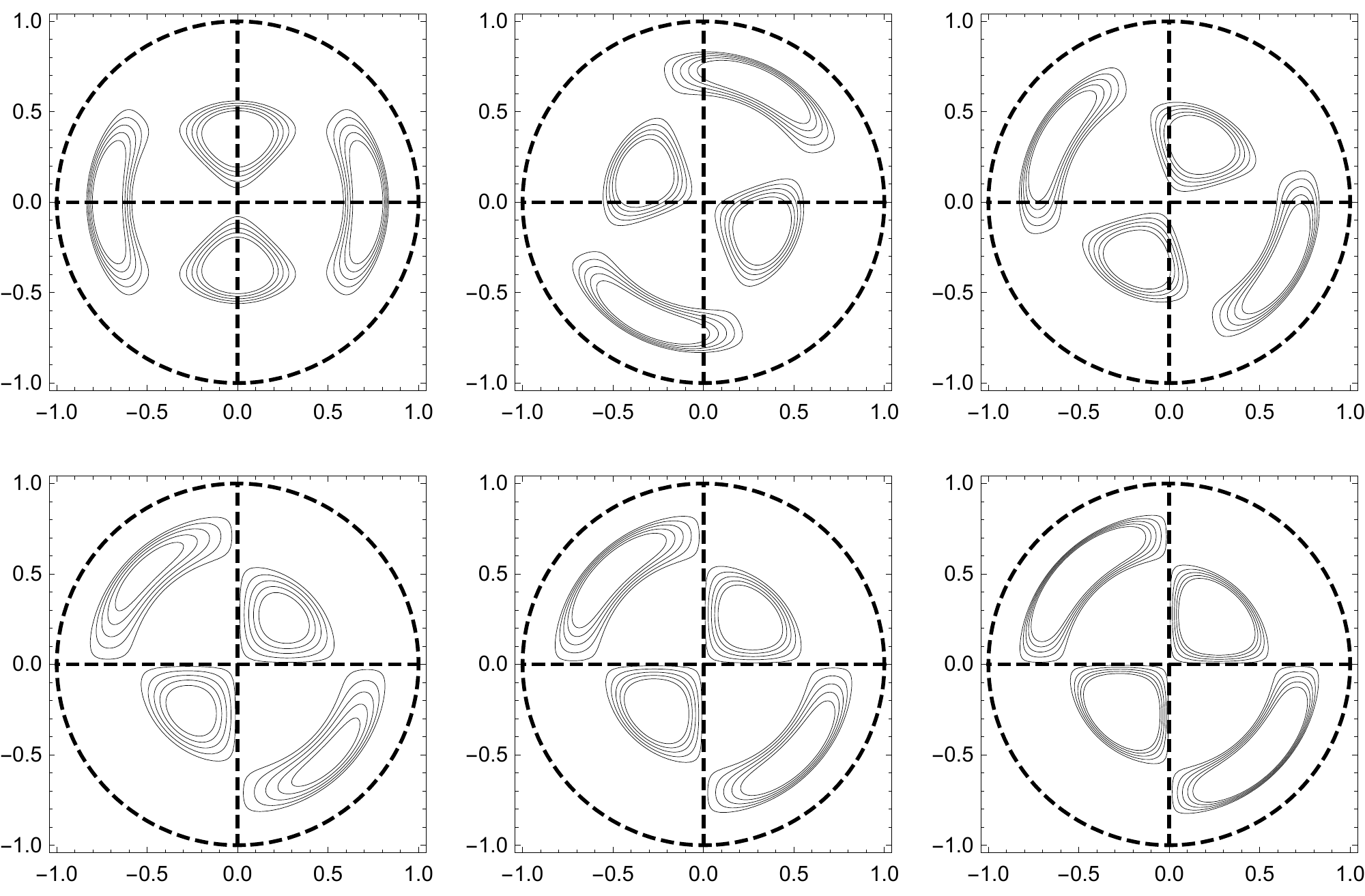}}
\caption{The contour plots show the real part of $\sigma_1$ for a $m=2$ mode. The top three plots belong to the fifth eigenmode of $m=2$ in Newtonia gravity with $\omega=-7.57$ and $\beta=1.1$. From left to right these three plots are in the time $t=0, 0.3,0.5$ respectively. The bottomn three plots belong to the same mode in MOG with $\alpha=8$, $\mu_0=0.01$ and $\beta=1.1$. In this case the eigenvalue is $\omega=-0.032 + 10.89$. From left to right these three plots are in the time $t=0, 0.03,0.07$ respectively. \label{fig4}}
\end{figure*}
As we mentioned before, the current observational constraints on the MOG's free parameters are $\alpha=8.89\pm 0.34$ and $\mu_0=0.042\pm 0.004 kpc^{-1}$. These bounds have been obtained using the rotation curve data of spiral galaxies. It is important to stress here that these parameters are not universal. More specifically, they are functions of the physical properties of the astrophysical system and can vary from system to system. Consequently, the above values are, in the best situation, true for spiral galaxies. We recall that $\mu_0$ appears as the vector field mass in the MOG field equations. Therefore its dependency to the environment's physical properties is reminiscent of the screening effects in scalar tensor theories of gravity, see \citet{kh}. However such effects have been not investigated in MOG. 
 
Regarding the main approximation that we have used in this paper,  i.e. $\mu_0 a \ll 1$, our analysis make sense for galaxy models with "baryonic" radius smaller than $24 kpc$. This radius is large enough to encompass a wide range of spiral galaxies. In other words, our assumption in this paper is not too restrictive. On the other hand regarding the magnitude of $\alpha$, the Maclaurin disk will be strongly unstable against non-axisymmetric perturbations. We emphasis again that this disk has also many unstable modes in Newtonian gravity. However all of them can be avoided by enhancing the pressure support as $\beta>1$. However in MOG, not only there are many unstable modes, but some of them cannot be avoided. 

In Fig. \ref{fig2}b), we have shown the zoomed eigenvalue spectrum for $m=2$. The black points are the eigenvalues in Newtonian gravity with $\beta=0.1$. Since $\beta<1$ then we see that there are 10 unstable modes with $\omega_I\neq 0$ and 23 stable modes. The red points are the corresponding eigenvalues in MOG with $\mu_0=0.01$, $\alpha=8$ and $\beta=0.1$. In this case there are 22 unstable modes and 11 stable modes. Therefore as we mentioned before, MOG excites several instabilities. Also the spectrum in MOG is more compact around $\omega_R=0$ and more extended along the $\omega_R$ axis. This means that MOG reduces the oscillation frequency of the stable and overstable modes and on the other hand increases the growth rate of the unstable modes.

In Fig. \ref{fig3}, we have shown the propagation of a $m=4$ mode in Newtonian gravity and in MOG. It should be noted that the matrix $\mathbf{B}$ as a $N\times N$ matrix, has in principle $N$ eigenvalues/eigenvectors for each $m$. As we mentioned before, we choose $N=33$. In Fig. \ref{fig3} we have plotted the surface density perturbation corresponding to the fifth eigenvalue. In Newtonian gravity with $\beta=1.1$ the above mentioned dimensionless eigenvalue is $\omega=-9.07$. Therefore since $\omega_I=0$ this mode is stable and as it is clear from the top three plots in Fig. \ref{fig3} this perturbation rotates counterclockwise without any damping or amplification.

On the other hand, in bottom three contour plots in Fig. \ref{fig3}, the time evolution of the same mode is plotted in MOG. In this case $\alpha=8$, $\mu_0=0.01$ and $\beta=1.1$. As we mentioned before, MOG strongly changes the eigenvalue spectrum. In this case the eigenvalue is $\omega=0.023+ 10.82i$. Therefore, as expected the angular frequency is substantially reduced and the rotational direction is reversed. It is clear form Fig. \ref{fig3} that in a small fraction of time the perturbation is strongly amplified. In fact counter curves get closer with time, and the density of the counter curves increases. This means that the surface density amplifies, and we call it gravitational instability. Since the real part of the frequency is small relative to the imaginary part, then mode is strongly unstable while its rotation rate is small and can not be seen in Fig. \ref{fig3}.

Since, observationally, the bar mode is the most interesting mode in spiral galaxies, we performed a similar analysis for $m=2$. The result is shown in Fig. \ref{fig4}. Similar to the $m=4$ case, assuming the same value for $\beta$ for Newtonian and MOG disks, this mode is strongly amplified in MOG in a short time scale. However in Newtonian case, $m=2$ mode is stable and propagates smoothly.

As a final remark, we mention that the main result of this paper is somehow consistent with our recent results considering the local stability of the disks, \citet{ro2015}. In fact in \citet{ro2015} we have found the generalized version of the Toomre criterion in MOG and shown that MOG increases the growth rate of the local perturbations. In other words, galactic disk are more unstable against local axi-symmetric perturbations in MOG than in Newtonian gravity. On the other hand in the current paper we have reached to the conclusion that MOG also has destabilizing effects on the global stability of the Maclaurin disk.
\section{conclusion}
\label{con}
To shed light on the effects of modified gravity (MOG) on the evolution of the stellar bars in galaxtic disks, we studied the modal properties of the Maclaurin disk. This disk has an analytic eigenvalue spectrum in Newtonian gravity and can help us achieve a better understanding of the global stability and bar formation in real galaxies. We used a semi-analytic method to investigate the normal modes of the disk in MOG. 

We found that the disk is strongly unstable in MOG. More specifically, increasing MOG's free parameters the grow rate of the instability increases. Therefore both parameters have destabilizing effects on the Maclaurin disk. Interestingly, $\mu_0$ only destabilizes the bar mode. Albeit the disk is more sensitive to the changes in $\alpha$ than in $\mu_0$. When $\alpha<1$ then one can stabilize the disk by enhancing the pressure support of the disk. On the other hand, surprisingly, if $\alpha>1$ then the instability is not avoidable and increasing the pressure does not help to stabilize the disk. Albeit it is important to mention that one can not certainly claim that increasing the pressure support can not suppress the instability. In fact as we have already mentioned, our main field equations are not valid in high pressure disks, and we have limited our analysis to the weak field limit. In order to find a more reliable result for the effect of pressure on the stability of the disk, it seems necessary to add post-Newtonian corrections to the field equations and repeat the stability analysis.

By increasing $\alpha$ unstable modes with larger $m$ is excited and consequently disk will undergo more unavoidable unstable modes. 

As mentioned before, existence of a dark matter halo will stabilize the disk. One may expect MOG to have a similar effect. This is a natural expectation for modified theories which try to address the dark matter problem without using dark matter particles. For instance \citet{brada} have shown that disk galaxies are more stable in MOND than in Newtonian gravity. Also \citet{ti} have shown than MOND leads to weaker stellar bars than Newtonian gravity. However, it is not the case for MOG at least for the Macluarin disk. In other words, although a dark matter halo stabilize the disk, MOG strongly destabilize it. Obviously one can not conclude that MOG will destabilize other real galactic disk models. In fact it is needed to investigate more realistic models including the stellar components, the thickness of the disk, the bulge and other important features of a galactic disk. 
\acknowledgments
We thank the anonymous referee for useful and constructive comments.
\section{appendix}
Using the transformation $\eta=\sqrt{1-R^2/a^2}$ and the oblate
spheroidal coordinate system, the solution of the Poisson
equation (\ref{var}), at $z = 0$, can be written as, for more details
see \citet{hu},
\begin{equation}
\Psi(\eta,\varphi)=\sum_{l,m}^{\infty}A_{lm}P_{m+2l}^m(\eta)e^{i m\varphi}
\label{pott1}
\end{equation}
\begin{equation}
\Sigma(\eta,\varphi)=-\frac{2}{\pi^2(1+\alpha)G a}\frac{1}{\eta}\sum_{l,m}^{\infty}A_{lm}\frac{P_{m+2l}^m(\eta)}{g_{m+2l,m}}e^{i m\varphi}
\label{dens2}
\end{equation}
where $P_l^m(\eta)$ are associated Legendre polynomials of the first kind and $l$ and
$m$ are integers which $l-m$ is even. $A_{lm}$ are expansion coefficients. Also the coefficients $g_{m,l}$
are given by
\begin{equation}
g_{m,l}=-\frac{4}{\pi}\frac{q_l^{m}(0)}{q_{l}^{'m}(0)}=\frac{(l+m)!(l-m)!}{2^{2l-1}\left[\left(\frac{l+m}{2}\right)!\left(\frac{l-m}{2}\right)!\right]^2}
\end{equation}
our main purpose in this appendix is to solve the equation (\ref{vec}) for a flattened density $\rho=\Sigma(R) \delta(z)$. We remind that the scalar wave equation (\ref{vec}) is separable in the oblate spheroidal coordinate system, see \citet{li}. In this coordinate system with coordinate variables ($\xi$, $\eta$,$\varphi$) defined in (\ref{coo}), and by using the separation of variables as $\Phi= f(\varphi)H(\eta)X(\xi)$, one may simply find the following differential equations in the vacuum
\begin{equation}
\frac{d}{d\xi}\left[(1+\xi^2)\frac{dX}{d\xi}\right]-\left(\lambda_{mn}+c^2\xi^2-\frac{m^2}{1+\xi^2}\right)X=0
\label{27}
\end{equation}
\begin{equation}
\frac{d}{d\eta}\left[(1-\eta^2)\frac{dH}{d\eta}\right]+\left(\lambda_{mn}-c^2\eta^2-\frac{m^2}{1-\eta^2}\right)H=0
\label{28}
\end{equation}
\begin{equation}
\frac{d^2 f}{d\varphi^2}=-m^2 f
\end{equation}
where $\lambda_{mn}$ and $m$ are separation constants (or eigenvalues) and $c=\mu_0 a<1$. equations (\ref{27}) and (\ref{28}) are differential equations for radial ($X$) and angular ($H$) oblate spheroidal functions. Therefore the solution for $H$ is the spheroidal angular harmonics of the fist and second kinds $S^{(k)}_{mn}(c,\eta)$, $k=1,2$, defined as
\begin{eqnarray}
S_{mn}^{(1)}(c,\eta)&=&\sum_{r=0}^{\infty} d_{mn}^r(c)P_{m+r}^m(\eta)\\ \nonumber
S_{mn}^{(2)}(c,\eta)&=&\sum_{r=-\infty}^{\infty}d_{mn}^r(c)Q_{m+r}^m(\eta)
\end{eqnarray}
where summations are over even values of $r$ when $n-m$ is even, and over only odd values of $r$ when $n-m$ are odd. Also $Q_{m+r}^m(\eta)$ (with range $|\eta|>1$) are the associated Legendre functions of the second kind and $d_{mn}^r(c)$ are known coefficients, see \citet{li} for more details. On the other hand it is clear that the differential equation of $X$ can be converted to that of $H$ by using a new variable $i\xi$. Therefore one may express the solution for $X$ as $S^{(1)}_{mn}(c,i\xi)$. Since $-1<\eta<1$, $H$ can be written only by $S_{mn}^{(1)}$. Finally the solutions are
\begin{eqnarray}
H(\eta)&\propto & S_{mn}^{(1)}(c,\eta)\\ \nonumber
X(\xi) &\propto & S^{(1)}_{mn}(c,i\xi)\\ \nonumber
f(\varphi)&\propto & e^{i m \varphi}
\end{eqnarray}
Therefore the potential can be expanded whit respect to the following eigenfunctions
\begin{equation}
\phi_{mn}=B_{mn}S_{mn}^{(1)}(c,\eta)\frac{S_{mn}^{(1)}(c,i \xi)}{S_{mn}^{(1)}(c,0)}e^{i m \varphi}
\label{pmn}
\end{equation} 
This potential is a vacuum solution and so satisfies the equation (\ref{vec}) everywhere expect on the disk. It is needed to be an even function of $\xi$. Bearing in mind the definitions of the spheroidal wave functions, one may deduce that $n-m$ should be even integer. In this case the potential (\ref{pmn}) will be a continuous function across the disk. However because of the presence of the matter at $\xi=0$, the normal component of $\nabla \Phi_{mn}$ is not continuous. Integrating the differential equation (\ref{vec}) across the disk, one may find the surface density that generates the potential (\ref{pmn}) as 
\begin{equation}
\Sigma_{mn}=-\frac{1}{2\pi G \alpha}\left(\frac{1}{h_{\xi}}\frac{\partial \phi_{mn}}{\partial \xi}\right)_{\xi=0}
\end{equation}
where $h_{\xi}$ is a scalar factor related to the metric of the flat three dimensional Euclidean space in the spheroidal coordinate system as $h_{\xi}=\sqrt{g_{\xi\xi}}=a \eta$. Therefore one may straightforwardly obtain the surface density as
\begin{equation}
\Sigma_{mn}=\frac{2}{\pi^2 \alpha G a}\frac{1}{\eta}\frac{B_{mn}}{\gamma_{mn}}S_{mn}^{(1)}(c,\eta)e^{i m \varphi}
\label{dens1}
\end{equation}
where
\begin{equation}
\gamma_{m,n}=-\frac{4}{\pi}\frac{S_{mn}^{(1)}(c,0)}{S_{mn}'^{(1)}(c,0)}
\label{dens}
\end{equation}
where prime denotes derivative with respect to $\xi$. Equations (\ref{pmn}) and (\ref{dens}) are exact and we have not yet used the approximation $c=\mu_0 a \ll 1$. Now let us apply this approximation. Fortunately, spheroidal functions $S_{mn}^{(1)}(c,x)$ can be expanded as a power series in $c$ as follows
\begin{eqnarray}\label{bast}
S_{mn}^{(1)}(c,x)=P_n^m(x)+&c^2[\psi_{mn}P_{n-2}^m(x) +\psi_{mn}'P_{n+2}^m(x)]~~~~~~\\\nonumber +O(c^4)
\end{eqnarray}
where
\begin{eqnarray}
\psi_{m,n}&=&\frac{(n+m)(1-m-n)}{2(1+2n)(2n-1)^2}
\\ \nonumber \psi_{m,n}'&=&\frac{(1-m+n)(2-m+n)}{2(1+2n)(3+2n)^2}
\end{eqnarray}
We mention that one can easily use Mathematica to calculate this functions and their eigenvalues. One can check also the validity of (\ref{bast}) using this software. When $x=i \xi$ we will have $P_n^m(i\xi)$ which become large at large $\xi$. Therefore we have to use another independent Legendre associate functions $q_{n}^m$ instead of $P_n^m$, for more details see \citet{hu}. In this case the coefficients $\gamma_{m,n}$ can be written as
\begin{eqnarray}
\gamma_{m,n}=g_{m,n}[1+\psi_{mn} a^2\mu_0^2\left(\frac{q_{n-2}^m}{q_{n}^m}-\frac{q_{n-2}'^m}{q_{n}'^{m}}\right)\\\nonumber + \psi_{mn}' a^2\mu_0^2\left(\frac{q_{n+2}^m}{q_{n}^m}-\frac{q_{n+2}'^m}{q_{n}'^{m}}\right)]_{\xi=0}
\end{eqnarray}
since functions $q_n^m(\xi)$ and $q_{n}'^{m}(\xi)$ are known at $\xi=0$, see \citet{hu}, we can simplify $\gamma_{m,n}$ as follows
\begin{equation}
\gamma_{m,n}=g_{m,n}+\frac{a^2\mu_0^2}{\pi}\frac{2\Gamma\left(\frac{n-m+1}{2}\right)\Gamma\left(\frac{n+m+1}{2}\right)(-1)^{n-m}}{(4n^2+4n-3)\Gamma\left(\frac{n-m+2}{2}\right)\Gamma\left(\frac{n+m+2}{2}\right)}
\label{fgamma}
\end{equation}
where $\Gamma$ is the Gamma function. As mentioned before, our aim in this paper is to study the stability of non-axisymmetric normal modes ($m\neq 0$). Therefore the general solution for the surface density, for a fixed mode $m$, can be expressed in terms of the associate Legendre functions as 
\begin{eqnarray}\label{f1}
\Sigma(\eta)=\frac{2}{\pi^2 \alpha G a}\frac{1}{\eta}\sum_{l=0}^{\infty}\frac{B_{l}e^{i m \phi}}{\gamma_{m,m+2l}}\Big[P_{m+2l}^m(\eta)~~~~~~~~\\  \nonumber\\ \nonumber +a^2\mu_0^2\left(\psi_{m,m+2l}P_{m+2l-2}^m(\eta)+\psi_{m,m+2l}'P_{m+2l+2}^m(\eta)\right)\Big]
\end{eqnarray}
where we have used (\ref{dens1}) and (\ref{bast}). Also since $n-m$ is even we have assumed $n=m+2l$. In the above equation, the surface density is expanded using the eigen functions of the Helmholtz equation (\ref{vec}). On the other hand, we can express it in terms of the eigen functions of the Poisson equation (\ref{var}). In other words, using equation (\ref{dens2}), we obtain the following expansion
\begin{eqnarray}\label{dj}
\Sigma(\eta)=-\frac{2}{\pi^2 (1+\alpha) G a}\frac{1}{\eta}\sum_{l=0}^{\infty}\frac{A_{l}e^{i m \phi}}{g_{m,m+2l}}P_{m+2l}^m(\eta)
\end{eqnarray}
Equating equations (\ref{f1}) and (\ref{dj}), helps to find a relation between coefficients $a_l$ and $b_l$. In fact, we multiply these equations with $P_{m+2k}^m(\eta)$ and integrate over interval $(0<\eta<1)$. Using the orthogonality condition of the associate Legendre functions, we find 
\begin{equation}
A_k=\alpha_k B_k+a^2\mu_0^2(\epsilon_k B_{k+1}+\epsilon_k' B_{k-1})
\label{rel}
\end{equation}
it is easy to show that $B_{-1}=0$. Also for brevity in notation we have defined the following new parameters
\begin{eqnarray}\label{ak}
\alpha_k &=&-\frac{1+\alpha}{\alpha}\frac{g_{m,m+2k}}{\gamma_{m,m+2k+2}}\\ \nonumber \\ \nonumber
\epsilon_k &=&-\frac{1+\alpha}{\alpha} \frac{\psi_{m,m+2k+2}g_{m,m+2k}}{\gamma_{m,m+2k+2}} \\ \nonumber
\epsilon_k' &=&-\frac{1+\alpha}{\alpha} \frac{\psi_{m,m+2k-2}'g_{m,m+2k}}{\gamma_{m,m+2k-2}}
\end{eqnarray}
Equation (\ref{rel}) gives $A$ coefficients with respect to $B$ coefficients. It is also useful to find the inverse relation. To do so let us rewrite equation (\ref{rel}) as follows
\begin{eqnarray}\label{rel2}
B_k=\frac{A_k}{\alpha_k}-\left(\frac{\epsilon_k}{\alpha_k}B_{k+1}+\frac{\epsilon_k'}{\alpha_k}B_{k-1}\right)a^2\mu_0^2
\end{eqnarray}
then one may write equation (\ref{rel2}) for $k+1$ and substitute $B_k$ and $B_{k+2}$ from equation (\ref{rel2}). Neglecting terms containing higher orders of $a^2\mu_0^2$, we find
\begin{equation}
B_{k+1}=\frac{A_{k+1}}{\alpha_{k+1}}-\left(\frac{\epsilon_{k+1}A_{k+2}}{\alpha_{k+1}\alpha_{k+2}}+\frac{\epsilon_{k+1}'A_{k}}{\alpha_k\alpha_{k+1}}\right)a^2\mu_0^2
\label{rel3}
\end{equation}
Before closing this section we mention that the effective potential $\Phi(\eta)$ on the surface of the disk can be expressed as
\begin{eqnarray}\label{fpot}
\Phi(\eta)=\sum_{l=0}^{\infty}(A_l+B_l)P_{m+2l}^m(\eta)+a^2\mu_0^2 B_l[\psi_{m,m+2l}\\\nonumber P_{m+2l-2}^m(\eta)+\psi_{m,m+2l}'P_{m+2l+2}^m(\eta)]
\end{eqnarray}
where we have used equations (\ref{pot1}), (\ref{pott1}), (\ref{pmn}) and (\ref{bast}).
\bibliographystyle{mn2e} 
\bibliography{manuscript}

\begin{thebibliography}{25}
\expandafter\ifx\csname natexlab\endcsname\relax\def\natexlab#1{#1}\fi
\bibitem[{{Athanassoula} (2002)}]{at2002} {Athanassoula}, E. 2002, \apj, 569, L83 
\bibitem[{{Athanassoula} \& {Selwood} (1986)}]{at1986} {Athanassoula}, E. \&  {Sellwood}, J. A. 1986, \mnras, 221, 213
\bibitem[{{Bekenstein} (2004)}]{bek} {Bekenstein}, J. D. 2004, PhRvD, 70, 083509
\bibitem[{{Binney} \& {Tremaine} (2008)}]{bt} {Binney}, J. \& {Tremaine}, S. 2008, Galactic Dynamics (2nd ed.; P
rinceton, NJ: Princeton Univ. Press)
\bibitem[{{Brada} \& {Milgrom} (1999)}]{brada} {Brada}, R., \& {Milgrom}, M. 1999, \apj, 519, 590
\bibitem[{{Brandao} \& {de Araujo} (2010)}]{brand} {Brandao}, C. S. S., \& {de Araujo}, J. C. N. 2010, \apj, 717, 849
\bibitem[{{Brownstein} \& {Moffat} (2006)}]{br2006} Brownstein J. R., Moffat J. W., 2006, \apj, 636, 721
\bibitem[{{Brownstein} \& {Moffat} (2007)}]{br2007} Brownstein J. R., Moffat J. W., 2007, \mnras, 382, 29
\bibitem[{{Buta} \& {Combes} (1996)}]{com} Buta, R. \& Combes, F. 1996, Fundamentals of Cosmic Physics, 17, 95
\bibitem[{{Capozziello} \& {De Laurentis} (2011)}]{ca} {Capozziello}, S. \& {De Laurentis}, M. 2011, PhR, 509, 167 
\bibitem[{{Christodoulou} (1991)}]{chris} Christodoulou, D. M. 1991, \apj, 372, 471
\bibitem[{{Efstathiou} et al (1982)}]{ef} Efstathiou G., Lake, G., Negroponte, J. 1982, \mnras, 199, 1069
\bibitem[{{Evans} \& {Read} (1998)}]{ev} Evans, N. W., \& Read, J. C. A. 1998, \mnras, 300, 106
\bibitem[{{Famaey} \& {McGaugh} (2012)}]{fa} Famaey, B. \& McGaugh, S.S. 2012, LRR, 15, 10
\bibitem[{{Gradshteyn} \& {Ryzhik} (2007)}]{gr} Gradshteyn, I.S. \& Ryzhik, I.M., 2007, Table of Integrals, Series, and Products
\bibitem[{{Hohl} (1971)}]{ho} Hohl, F., 1971, \apj, 168, 343.
\bibitem[{{Hunter} (1963)}]{hu} Hunter, C. 1963, \mnras, 126, 299
\bibitem[{{Jalali} (2007)}]{jal2007} Jalali M. A., 2007, \apj, 669, 218
\bibitem[{{Jalali} \& {Hunter} (2005)}]{jal2005} Jalali M. A. \& Hunter C., 2005, \apj, 630, 80
\bibitem[{{Jamali} \& {Roshan} (2016)}]{Jamali} Jamali S. \& Roshan M., 2016, EPJC, 76, no. 9, 490; arXiv:1608.06251 
\bibitem[{{Kalnajs} (1972)}]{ka} Kalnajs, A. J. 1972, \apj, 175, 63
\bibitem[{{Khoury} \& {Weltman} (2004)}]{kh} Khoury, J. \& Weltman, A. 2004, PhRvD, 69, 044026
\bibitem[{{Li} et al (2002)}]{li} Li et al, 2002, Spheroidal wave functions in electromagnetic theory, (New York, John Wiley \& Sons)
\bibitem[{{Martinez-Valpuesta} et al (2006)}]{ma} Martinez-Valpuesta, I., et al. 2006, \apj, 637, 214
\bibitem[{{Milgrom} (1983)}]{milgrom} Milgrom, M. 1983, \apj, 270, 384
\bibitem[{{Miller} et al (1970)}]{miller} Miller, R. H., Prendergast, K. H., Quirk, W. J. 1970. \apj, 161, 903
\bibitem[Moffat (2006)]{mo2006} Moffat J. W., 2006, JCAP, 0603, 004
\bibitem[Moffat (2015)]{mo2015} Moffat J. W., 2015, EPJC, 75, no. 3, 130 
\bibitem[Moffat \& Rahvar (2013)]{mo2013} Moffat J. W., Rahvar S., 2013, \mnras, 436, 1439
\bibitem[Moffat \& Rahvar (2014)]{mo2014} Moffat J. W., Rahvar S., 2014, \mnras, 441, 3724
\bibitem[Moffat \& Toth (2008)]{mo2008} Moffat J. W., Toth V. T., 2008, \apj, 680, 1158
\bibitem[Moffat \& Toth (2009)]{mo2009} Moffat J. W., Toth V. T., 2009, CQGra, 26, 085002
\bibitem[Moffat \& Toth (2013)]{mo2013a} Moffat J. W., Toth V. T., 2013, Galax, 1, 65
\bibitem[Ostriker \& Peebles (1973)]{os} Ostriker, J. P. \& P. J. E. Peebles, 1973, \apj, 186, 467
\bibitem[Poisson \& Will (2014)]{po} Poisson, E. \& Will, C. 2014, Gravity: Newtonian, Post-Newtonian, Relativistic (New York, Cambridge university press)
\bibitem[Polyachenko (1977)]{polyachenko} Polyachenko, V.L. 1977, SvAL, 3, 51
\bibitem[Roshan \& Abbassi (2015)]{ro2015} Roshan, M., \& Abbassi, S. 2015, \apj, 802, no. 1, 9; arXiv:1501.04715
\bibitem[Roshan \& Abbassi (2014)]{ro2014} Roshan, M., \& Abbassi, S. 2014, PhRvD, 90, no. 4, 044010; arXiv:1407.6431
\bibitem[Roshan (2013)]{ro2013} Roshan, M., 2013, PhRvD, 87, 044005; arXiv:1210.3136  
\bibitem[Roshan (2015)]{roepjc} Roshan, M., 2015, EPJC, 75, no. 9, 405; arXiv:1508.04243 
\bibitem[Saha \& Naab (2013)]{saha} Saha K., Naab T., 2013, \mnras, 434, 1287
\bibitem[Sanders (2010)]{sand2010} Sanders R. H. , 2010, the dark matter problem: 
A Historical Perspective (Cambridge University Press, Cambridge).
\bibitem[{{Sanders} \& {Huntley} (1976)}]{san1976} Sanders, R. H. \& Huntley, J. M. 1976, \apj, 209, 53
\bibitem[{{Sellwood} (1981)}]{se} Sellwood, J. A., 1981, \aap. 99, 362
\bibitem[{{Sellwood} (2016)}]{se2016} Sellwood, J. A., 2016, \apj. 819, no. 2, 92
\bibitem[{{Sheth} et al (2008)}]{sh} Sheth, K. et al. 2008, \apj, 675, 1141
\bibitem[{{Takahara} (1976)}]{ta1976} Takahara, F. 1976, PThPh, 56, 1665
\bibitem[{{Takahara} (1978)}]{ta1978} Takahara, F. 1978, PASJ, 30, 253
\bibitem[{{Tiret} \& {Combes} (2007)}]{ti} Tiret, O., \& Combes, F. 2007, \aap, 464, 517
\bibitem[{{Toomre} (1969)}]{to} Toomre, A. 1969, \apj, 158, 899
\bibitem[{{Weinberg} (1983)}]{we} Weinberg, M.D. 1983, \apj, 271, 595
\end{thebibliography}
\label{lastpage}
\end{document}